\documentstyle[12pt,a4,psfig]{article}
\newcommand{\be}         {\begin{equation}}
\newcommand{\bea}        {\begin{eqnarray}}
\newcommand{\cf}         {{\it cf.\,}}
\newcommand{\ee}         {\end{equation}}
\newcommand{\eea}        {\end{eqnarray}}
\newcommand{\eg}         {{\it e.g.\,}}
\newcommand{\eps}        {\varepsilon}
\newcommand{\eqn}[1]     {eq.\,(\ref{#1})}
\newcommand{\expv}[1]    {\left\langle#1\right\rangle}
\newcommand{\fm}         {\mbox{\,fm}}
\newcommand{\gapp}       {\lower.7ex\hbox{$\;\stackrel{\textstyle>}{\sim}\;$}}

\newcommand{\lapp}       {\lower.7ex\hbox{$\;\stackrel{\textstyle<}{\sim}\;$}}
\newcommand{\MeV}        {\mbox{\,MeV}}
\newcommand{\rmi}        {{\rm{i}}}
\newcommand{\bmp}[2]     {\begin{minipage}[#1]{#2}} 
\newcommand{\emp}        {\end{minipage}}
\newcommand{\cE}{{\cal E}}

\begin{document}                                                               
\pagestyle{empty}
\begin{center}\large
\bf{Self-Consistent Pushing and Cranking Corrections \\ 
to the Meson Fields of the Chiral Quark-Loop Soliton}
\end{center}\normalsize
\vskip\baselineskip\vskip\baselineskip
\begin{center}
{\sc M.\,Schleif$^1$, R.\,W\"unsch$^1$ and T.\,Meissner$^2$}
\end{center}
\bigskip \bigskip
$^1$ Institut f\"ur Kern- und Hadronenphysik, Forschungszentrum 
Rossendorf e.\,V.,\\ 
\hspace*{3mm}Postfach 51\,01\,19, D-01314 Dresden, Germany

\noindent
$^2$ Department of Physics, Carnegie-Mellon-University, Pittsburgh, PA 15213,\\
\hspace*{3mm}USA

\vskip\baselineskip
\vskip\baselineskip
\vskip\baselineskip

\begin{abstract}
We study translational and spin-isospin symmetry restoration for the 
two-flavor chiral quark-loop soliton. 
Instead of a static soliton at rest we consider a boosted and rotating 
hedgehog soliton. 
Corrected classical meson fields are obtained by minimizing a corrected energy 
functional which has been derived by semi-classical methods
('{\it variation after projection}'). 

We evaluate corrected meson fields in the region 
$300\MeV\!\le\!M\!\le\!600\MeV$
of constituent quark masses $M$ and compare them with the uncorrected fields.
We study the effect of the corrections on various expectation values of 
nuclear observables such as the root-mean square radius, the 
axial-vector coupling 
constant, magnetic moments and the delta-nucleon mass splitting.
\end{abstract}
\newpage

\setcounter{page}{1}
\pagestyle{plain}

\section{Introduction}
Chiral soliton models have turned out to be a fruitful approach to the
structure of light baryons. We consider the soliton of the 
two-flavor chiral quark-loop model, which is equivalent to
the semi-bosonized Nambu \& Jona-Lasinio (NJL) model, with 
hedgehog meson fields
restricted to the chiral circle 
(for a detailed review \cf ref.\,\cite{Christov-96,Alkofer-96}). 
Mesons are treated as mean fields (zero boson loop) whereas the 
polarization of the Dirac sea is fully taken into account.
In order to obtain a system with baryon number $B$=1 one has to add 
$N_{\rm{c}}$=3 valence quarks occupying the lowest single-particle level
with positive energy.
Static, self-consistent mean-field solitons are obtained by minimizing the
effective energy.
For a detailed description of procedure, formalism and notation 
we refer the reader to ref.\,\cite{Wuensch-94}.

As it is well known the mean-field hedgehog soliton breaks both the 
translational and rotational symmetry. 
The soliton is not an eigenstate of the total linear 
($\vec{P}$) and angular ($\vec{J}$) momentum and of the isospin operator 
($\vec{T}$). 
As a result, the expectation value $\expv{\vec{P}^{\,2}}$ does not vanish, and 
$\expv{\vec{J}^{\,2}}$ and $\expv{\vec{T}^{\,2}}$ do not
have the values appropriate for a nucleon or a $\Delta$ isobar.
The energy functional, which defines the solitonic field configuration, is 
contaminated by spurious contributions.
There are various method to exclude the spurious energies and to determine
the energy of a state with definite quantum numbers.
In non-relativistic many-particle physics the spurious energies can be
obtained within a certain approximation using Peierls-Yoccoz 
\cite{Peierls-57} or Peierls-Thouless \cite{Peierls-62} projection
techniques for the linear and angular momenta. 
Unfortunately those methods require the definition of a Fock state which 
is very involved if dealing with the polarized and regularized Dirac sea. 
In the case of the Skyrme model it has turned out that the consideration of 
RPA fluctuations around the mean-field solution allows a treatment of the 
corresponding eigenmodes \cite{Moussalam-91,Holzwarth-92} analogously to the 
case of two-dimensional soliton models \cite{Rajaraman-82}. 
Recently this approach has also been used to calculate the quantum corrections 
to the effective soliton energy in the chiral quark-loop model within some 
approximations \cite{Weigel-94}.

In the present paper we follow a simplified approach and rely on the
semi-classical pushing and cranking approximations \cite{Ring-80}.
Hereby one is considering 
a soliton with the appropriate quantum numbers by pushing 
the static hedgehog with the velocity $V$ and rotating its 
isospin coordinates with the angular velocity $\Omega$ where $V$
and $\Omega$ are regarded as Lagrange multipliers.
Due to the hedgehog symmetry the angular momentum of the soliton is fixed to 
$\vec{J}$=$-\vec{T}$ and one has automatically
$\expv{\vec{J}^{\,2}}\!=\!\expv{\vec{T}^{\,2}}$.
All the studies which have been done so far performed
{\em the projection after the variation},
which means that first the mean field configurations
are calculated by minimizing the static soliton energy
and afterwards the rotational and translational correction terms are added.
It is the aim of this paper to investigate if and how the properties 
of the chiral quark-loop model change if the more accurate 
procedure of a {\em variation after projection} is carried out. 
In this case one has to minimize a soliton energy which includes the
correction term. The resulting {\em corrected meson fields} respond 
to the collective translation and rotation of the soliton 
and deviate from the static ones.
The inertial parameters correspond to those values which minimize the corrected
energy functional ({\em self-consistent cranking}).
Expectation values of quark and meson observables are also modified.

In sect.\,2, we define the static hedgehog soliton and introduce a corrected 
energy functional for a soliton with definite momentum and isospin. The
field configuration which minimizes the corrected energy functional is 
determined in sect.\,3. Particular attention is drawn to the asymptotic 
behavior of the meson profile.
In sect.\,4 we investigate the effect of the corrections on solitonic
expectation values.
Conclusions are drawn in sect.\,5.

\section{Static mean-field and energy corrections}
 
In mean-field approximation, the SU(2) NJL lagrangian describes $u$ and $d$
quarks interacting with classical meson fields $\sigma$ and $\hat{\vec{\pi}}$.
Restricted to time-independent spherical hedgehog configurations
\Big($\sigma(t,\vec{r})\!=\!\sigma(r)$, 
$\vec{\pi}(t,\vec{r})\!=\!\pi(r)\,\hat{\vec{r}}$\Big) and to the 
chiral circle \Big($\sigma(r)^2+\pi(r)^2\!=\!M^2$\Big) with unit vector
$\hat{\vec{r}}\!\equiv\!\vec{r}/r$, $r\!=\!|\vec{r}|$, the meson fields are 
uniquely determined by the constituent quark mass $M$ and the profile function
$\Theta(r)\!\equiv\!\arctan{\frac{\pi(r)}{\sigma(r)}}$. 
The mean-field energy
\be\label{Emf}
E_{\rm{mf}}[\Theta]=E^{\rm{m}}[\Theta]+E^{\rm{q}}[\Theta]
\ee
is a functional of $\Theta$ and consists of a purely 
mesonic part
\be\label{Em}
E^{\rm{m}}[\Theta]=\frac{m_0M}{G}
   4\pi\int\!r^2dr\,\left[1-\cos{\Theta(r)}\right],
\ee
and of the quark energy
\be\label{Eq}
E^{\rm{q}}[\Theta]=
-\lim_{T_\cE\to\infty}\frac{1}{T_{\cE}}{\rm{Tr}}\,
      {\rm{Ln}}\left[\partial_\tau+h\right]
+\lim_{T_\cE\to\infty}\frac{1}{T_{\cE}}{\rm{Tr}}\,
      {\rm{Ln}}\left[\partial_\tau+h_0\right]
+N_{\rm{c}}\,\eps_{\rm{val}}\,\Theta(\eps_{\rm{val}})
\ee
with the single-particle quark hamiltonians
\be\label{h}
h=-\rmi\vec{\alpha}\!\cdot\!\vec{\nabla}+
  \beta M e^{\rmi\gamma_5\hat{\vec{\tau}}
  \cdot\hat{\vec{r}}\,\Theta(r)}
\qquad\mbox{and}\qquad
h_0=-\rmi\vec{\alpha}\!\cdot\!\vec{\nabla}+\beta M 
\;.\ee 
Here, $\vec{\alpha}\!\equiv\!\beta\vec{\gamma}$, 
$\vec{\gamma}\!\equiv\!(\gamma^1,\gamma^2,\gamma^3)$,
$\gamma_5\!\equiv\!\rmi\gamma^0\gamma^1\gamma^2\gamma^3\gamma^4$ and 
$\beta\!\equiv\!\gamma^0$
are Dirac matrices, $m_0$ is an average current quark mass 
$m_0\!=\!(m_{\rm{u}}+m_{\rm{d}})/2$. 
The Euclidean time-coordinate is denoted by $\tau$, 
while $\hat{\vec{\tau}}$ denotes the vector of Pauli matrices.
The last term in eq.\,(\ref{Eq}) describes the energy of the valence quarks
which have been added to ensure a baryon number $B\!=\!1$ for the soliton.
It vanishes if the meson field is so strong that the energy $\eps_{\rm{val}}$
of the valence level is negative. 
In this case the Dirac sea has already $B\!=\!1$.
The trace $\rm{Tr}$ in eq.\,(\ref{Eq}), which includes functional trace with 
anti-periodic boundary conditions in the Euclidean time interval 
$(-\frac{T_{\cE}}{2},\frac{T_{\cE}}{2})$
and matrix trace over Dirac, flavor and color indices, can be expressed by a
regularized sum over the eigenvalues of the hamiltonians $h$ and $h_0$.
For details \cf ref.\,\cite{Wuensch-94}. 

The classical meson profiles have to minimize the mean-field energy 
(\ref{Emf})
\be\label{eom}
\frac{\delta E_{\rm{mf}}}{\delta\Theta(r)}=0
\ee
leading to the {\em equation of motion}
\be\label{eom1}
\Theta(r)=\arctan{\frac{\tilde{P}(r)}{\tilde{S}(r)}}
\;,\ee
which is an implicit equation since scalar and pseudoscalar quark
densities $\tilde{S}(r)$ and $\tilde{P}(r)$ (\cf ref.\,\cite{Wuensch-94}) 
are functionals of the meson fields.
At large separations from the center of the soliton the meson fields 
approach to their vacuum values $\sigma(r\!\to\!\infty)\!=\!\sigma_0\!=\!M$ 
and $\pi( r\!\to\!\infty)\!=\!0$, while the difference 
$\Theta(0)-\Theta(\infty)$ is related to the baryon number of the soliton. 
In order to obtain a soliton with baryon number $B\!=\!1$ we assume 
\be\label{ThAsy0}
\Theta(r)\longrightarrow\left\{
   \begin{array}{lll}-\pi& \mbox{for}&r\to0\\0&\mbox{for}&r\to\infty
   \end{array}\right.
\;.\ee
The implicit equation of motion (\ref{eom1}) is usually solved iteratively. 
Starting from a reasonable profile $\Theta^0$ one diagonalizes the 
hamiltonians (\ref{h}) and gets a set of eigenvalues and eigenfunctions 
which allow to determine the densities $\tilde{S}$ and $\tilde{P}$. 
The latter are used on the right side of eq.\,(\ref{eom1}) in order to get 
an improved profile. 
This procedure is repeated unless the profile remains unchanged within a 
certain accuracy. 
It has been shown \cite{Reinhardt-88} that the iteration converges for 
constituent quark masses $M\gapp330\MeV$. 
Mesonic fields which fulfill the equation of motion (\ref{eom1}) with quark 
fields which diagonalize the corresponding quark hamiltonian (\ref{h}) are 
called {\em self-consistent} solutions of the NJL model in mean-field 
approximation. 
They have to be distinguished from a parameterized meson field with a 
predetermined shape which approximates the self-consistent field and does 
not fulfill the equation of motion (\ref{eom1}) in general.

The soliton as a mean-field solution with hedgehog shape is not an eigenstate
of total momentum $\vec{P}$ and isospin operator $\vec{T}$.
As a result, the expectation values $\expv{\vec{P}^{\,2}}$ and 
$\expv{\vec{T}^{\,2}}$ do not have the values
\be\label{P2T2desired}
\expv{\vec{P}^{\,2}}=0 \qquad\mbox{and}\qquad 
\expv{\vec{T}^{\,2}}=T(T+1)
\ee
with $T\!=\!\frac{1}{2}$ for a nucleon and $T\!=\!\frac{3}{2}$ for a 
$\Delta$ isobar at rest.  
The mean field distinguishes a definite position (the center) of the soliton 
and the hedgehog defines a privileged direction ($\tau_z$ is maximal in $z$
direction) what obviously contradicts translational and rotational symmetry.
The mean-field hedgehog soliton performs spurious motions 
(oscillations around the center of mass and in isospace). 
These spurious motions are not only responsible for the deviation of 
$\expv{\vec{P}^{\,2}}$ and $\expv{\vec{T}^{\,2}}$ from the physical values 
(\ref{P2T2desired}) but contribute to most of the expectation values of 
the soliton.

The energy of the spurious modes can be approximately calculated in the 
framework of the pushing and cranking approaches often
used in non-relativistic many-particle physics \cite{Ring-80}.
After semi-classical quantization \cite{Adkins-83} of the rotational degrees 
of freedom the energy of a soliton with the expectation values 
(\ref{P2T2desired}) is given by \cite{Pobylitsa-92}
\be\label{Ecorr}
E^T_{\rm{corr}}=E_{\rm{mf}}-E^0_{\rm{trans}}-E^0_{\rm{rot}}+E^T_{\rm{crank}}
\ee
with the translational zero-point  (center-of-mass) energy
\be\label{E0trans}
E^0_{\rm{trans}}\,=\,\frac{\expv{\vec{P}^{\,2}}_{\rm{hh}}}{2M^{\rm{inert}}}
\ee 
and its iso-rotational equivalent
\be\label{E0rot}
E^0_{\rm{rot}}\,=\,\frac{\expv{\vec{T}^{\,2}}_{\rm{hh}}}{2I}\,=\,
\frac{9}{8I}
\;.\ee
The parameters $M^{\rm{inert}}$ and $I$ are the Thouless-Valatin 
inertial parameters of translational and rotational motion of the soliton 
as a whole.
As shown in \cite{Pobylitsa-92} the inertial mass $M^{\rm{inert}}$
coincides with the static mean-field energy (\ref{Emf}). The expectation 
value $\expv{\vec{P}^{\,2}}_{\rm{hh}}$ of the mean-field hedgehog soliton 
consists of contributions from  valence and sea quarks 
and is a functional of the 
meson profile. It diverges and therefore has to be regularized. 
We apply Schwinger's proper-time scheme \cite{Schwinger-51} and relate 
the cut-off parameter to the constituent quark mass by means of the 
weak pion decay. For details see 
refs.\,\cite{Reinhardt-88,Reinhardt-89,Reinhardt-89a,Meissner-91}.

The expectation value of 
$\expv{\vec{T}^{\,2}}_{\rm{hh}}$ is finite. 
In contrast to the corresponding value for the linear 
momentum it is independent of the shape of the profile function 
and consists of the single-particle expectation values of the 3 valence quarks 
$\expv{\vec{T}^{\,2}}_{\rm{hh}}\!=\!
3\cdot\frac{1}{2}\left(\frac{1}{2}+1\right)\!=\!\frac{9}{4}$. 
The Dirac sea does not contribute to the isospin of the hedgehog soliton.
The iso-rotational moment of inertia $I$ is given by a
regularized Inglis formula \cite{Reinhardt-89} and has to be calculated 
numerically.
A term like (\ref{E0rot}) has a well known analog 
in nuclear physics, namely the
band-head energy which has to be subtracted from the mean field energy
if the cranking procedure is applied to a nucleus with spin different 
from zero.
The last term in eq.\,(\ref{Ecorr})
\be\label{Ecrank}
E^T_{\rm{crank}}\,=\,\frac{T(T+1)}{2I}
\ee
is the genuine cranking term and
results from the global quantized rotation which was introduced 
to obtain the correct value for the isospin.
Both the rotational corrections (\ref{E0rot}) and (\ref{Ecrank}) 
can be considered as the result of a single cranking procedure in the course of
which the expectation value $\expv{\vec{T}^{\,2}}$ has been changed from 
$\frac{9}{4}$ (resulting from 3 valence quarks) to the physical value 
(\ref{P2T2desired}). The corresponding semi-classically quantized angular 
velocity is given by
\be\label{Omega}
\Omega=-\rmi\frac{\sqrt{T(T+1)-\expv{\vec{T}^{\,2}}}_{\rm{hh}}}{I}
\ee
where the second term in the numerator stems from the rotational zero-point
energy of the hedgehog while the factor $-\rmi$ appears because we are working
in Euclidean space-time.
It is worthwhile to mention that the same correction terms (\ref{E0rot}) 
and (\ref{Ecrank}) can be obtained \cite{Ring-80} within an approximate 
projection method using the Kamlah expansion \cite{Kamlah-68}.
The various contributions (\ref{E0trans}--\ref{Ecrank}) to the corrected 
soliton energy (\ref{Ecorr}) are of different order in $N_{\rm{c}}$, 
namely $E_{\rm{mf}}\!\sim\!N_{\rm{c}}$, $E^0_{\rm{trans,rot}}\!\sim\! 1$ and 
$E^T_{\rm{crank}}\!\sim\! 1/N_{\rm{c}}$, even though it will turn out that 
the cranking term can be numerically of the same order of magnitude as the 
zero-point energies. \vspace*{-5mm}

\noindent
\bmp{h}{9cm}\hspace*{-4mm}
\mbox{\psfig{file=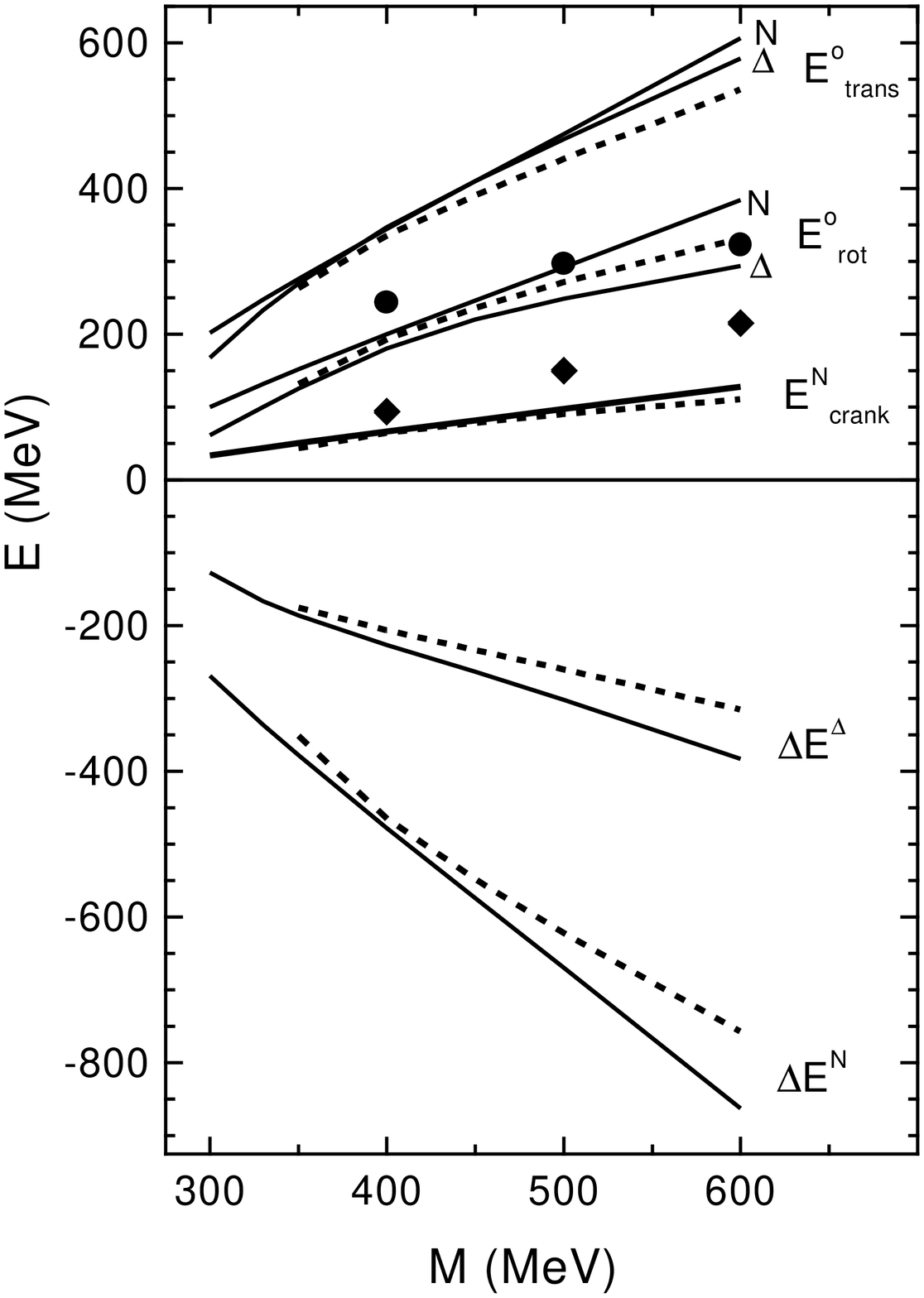,width=9cm,angle=0}}
\emp\hfill\bmp{h}{5.4cm}

{\small{\bf Fig.\,1:}\hspace*{1mm}
\mbox{Total energy corrections} 
$\Delta E^{\rm{N},\Delta}\!=\!-E^0_{\rm{trans}}\!-\!
E^0_{\rm{rot}}\!+\!E^{\rm{N},\Delta}_{\rm{crank}}$ for nucleon 
and $\Delta$ isobar ({\em lower part}) and their components ({\em upper part})
calculated for the static soliton profile ({\em broken lines}), 
and for the corrected nucleon ($N$) and $\Delta$ profiles ({\em full lines}) 
as a function of the constituent quark mass $M$. 
The cranking energy $E^\Delta_{\rm{crank}}$ of the $\Delta$ isobar is 
five times larger than the corresponding energy $E^{\rm{N}}_{\rm{crank}}$ 
of the nucleon.
The energies of the translational and rotational fluctuations calculated 
in reference \cite{Weigel-94} are indicated by $\Diamond$ 
and $\bullet$, respectively.\baselineskip10pt}
\emp

Fig.\,1 illustrates the size of the energy corrections. 
Each of the terms increases approximately linearly with the constituent 
quark mass $M$. The increase of the corrections is related to the spatial size
of the soliton. 
At larger constituent masses the quarks are stronger bound and the
soliton is smaller (\cf fig.\,7). 
Moment of inertia and expectation value $\expv{\vec{P}^{\,2}}_{\rm{hh}}$ 
behave accordingly. While the moment of inertia decreases the momentum
fluctuations increase if the size of the soliton is reduced at larger $M$.  
Already at small quark masses the corrections amount to a considerable part 
of the total mean-field soliton energy of roughly $1230\MeV$. 
At $M\gapp480\MeV$ the total energy correction for the nucleon exceeds 
one half of the soliton energy. 
For the $\Delta$ isobar the total energy correction is smaller since the 
zero-point corrections, which have to be subtracted, are partly compensated 
by the big positive cranking term.
Additionally we have indicated the energies of the corresponding mesonic 
quantum fluctuations calculated in RPA \cite{Weigel-94}.
Comparing them with the semi-classical zero-point energies (\ref{E0trans}) 
and (\ref{E0rot}) one has to take into account that only bound fluctuation
have been considered. 
According to the estimate of ref.\,\cite{Weigel-94} unbound fluctuations,
which appear in the non-confining NJL model, 
contribute roughly 50 percent to the translational modes, while unbound 
rotational modes are negligible.

The considerable size of the energy corrections in comparison to the
mean-field energy $E_{\rm{mf}}$ conflicts with their
perturbative calculation. To improve the approach we shall introduce 
corrected meson fields which minimize the corrected energy (\ref{Ecorr})
instead of the mean-field energy (\ref{Emf}).

\section{Corrected meson profiles}

In sect.\,2 we considered the static equation of motion (\ref{eom1})
and its solution $\Theta(r)$, in the following called ``uncorrected'' 
or ``static'' profile.
The energy corrections (\ref{E0trans}--\ref{Ecrank}) were calculated 
afterwards using the uncorrected profiles. This approach corresponds to a
{\em projection after variation}.  
In the present section we want to perform a {\em variation after projection} 
by minimizing the corrected energy \mbox{functional (\ref{Ecorr})} 
\be\label{eomcorr}
\left.\frac{\delta}{\delta\Theta(r)}\,E_{\rm{corr}}^T[\Theta]
\right|_{\Theta=\Theta^T_{\rm{corr}}}\,=\,0
\ee
from which we obtain ``corrected'' profiles 
$\Theta_{\rm{corr}}^T(r)$.
The corrected profiles deviate from the static ones because the inertial 
parameters $M^{\rm{inert}},I$ and the expectation value 
$\expv{\vec{P}^{\,2}}_{\rm{hh}}$ depend on the meson profile. 
Minimizing the corrected energy functional one allows them to acquire a value 
which corresponds to a lower corrected energy 
(self-consistent pushing and cranking). 
Since $E_{\rm{corr}}^T$ depends explicitly on the isospin quantum number
$T$ the corrected profiles are different for nucleons and $\Delta$ isobars.

The lower part of fig.\,2 illustrates the difference between the 
self-consistently determined moments of inertia for nucleons and $\Delta$ 
isobars and compares them with the moment of the static hedgehog soliton. 
Above $M\!\approx380\MeV$, the moment for the $\Delta$ 
isobar is roughly 15\% larger than for the nucleon. The harmonic average
between both momenta agrees quite well the static moment (\cf sect.\,4). 
At smaller constituent quark masses, the mean field has a shallow form and
the valence quarks rather weakly bound. 
In this case the valence quarks are allowed 
to travel to large distances from the center of the soliton where they are
strongly affected by the isospin dependent rotational terms in the 
corrected effective energy (\ref{Ecorr}).  The resulting moments of inertia 
are large and remarkably different for nucleon and $\Delta$ isobar. 
The root-mean square (r.\,m.\,s.)~radius shows a similar behavior (fig.\,7). 
At $M\lapp350\MeV$, the uncorrected mean field is too shallow to produce a
bound quark state and a self-consistent solitonic solution of the static
equation of motion (\ref{eom1}) does not exist. After subtracting the
translational zero-point energy (\ref{E0trans}) the soliton stabilizes and
a self-consistent field configuration exists up to $M\!\approx\!300\MeV$.
\vspace*{-5mm}

\noindent
\bmp{h}{9cm}\hspace*{-4mm}
\mbox{\psfig{file=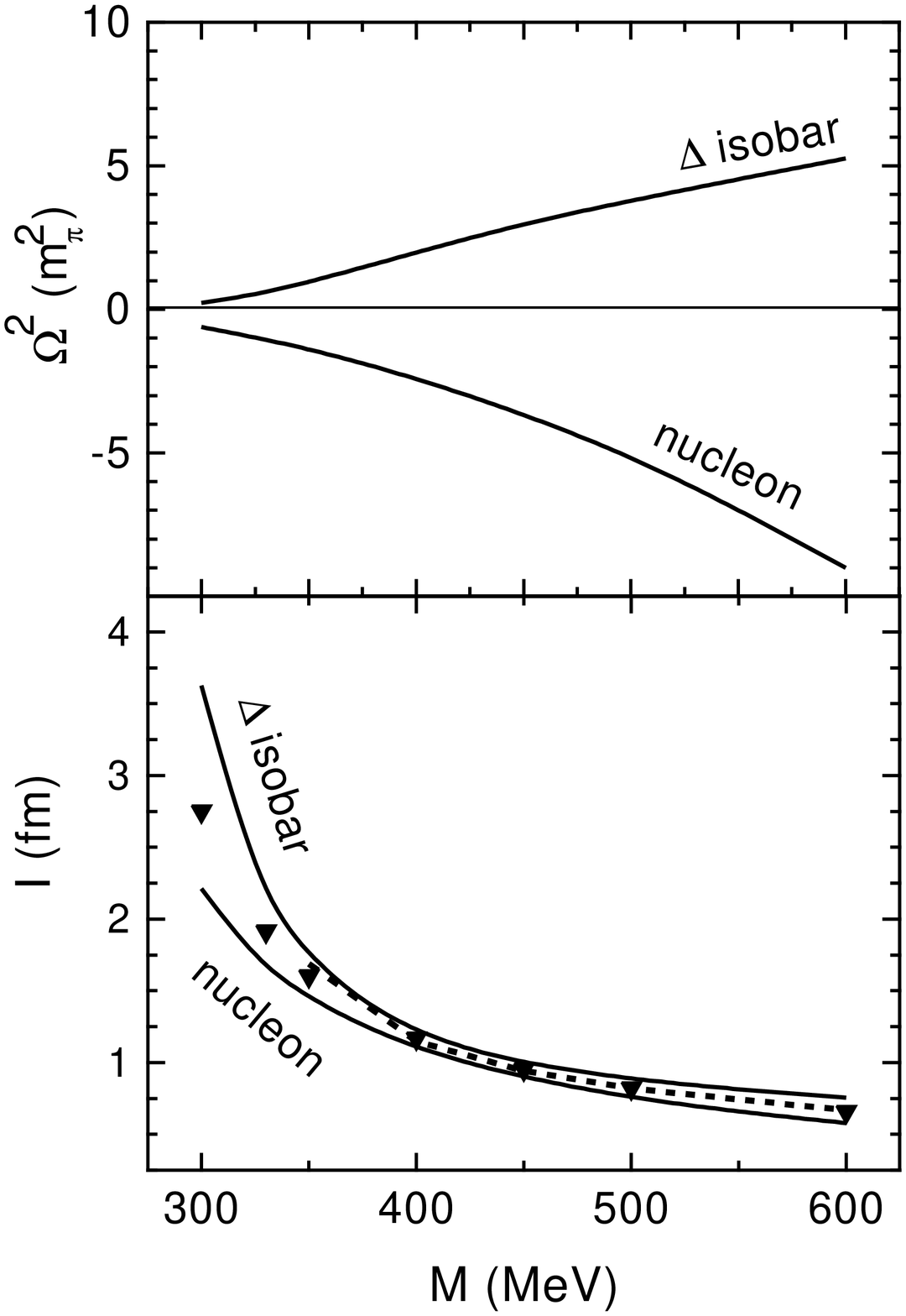,width=9cm,angle=0}}
\emp\hfill\bmp{h}{5.4cm}
{\small{\bf Fig.\,2:}\hspace*{1mm}
{\em Lower part}:\\ Self-consistently determined iso-rotational moments 
of inertia $I$ ({\em full lines}) of nucleon and $\Delta$ isobar 
in comparison to the static moment of the hedgehog soliton ({\em broken line})
as a function of the constituent quark mass $M$.
The triangles indicate the harmonic average of the self-consistent moments
of both particles.

{\em Upper part}: Lagrange parameter $\Omega^2$ according to 
eq.\,(\ref{Omega}) for nucleon and $\Delta$ isobar as a function of the
constituent quark mass.\baselineskip10pt}
\emp

The upper part of fig.\,2 shows the Lagrange parameter $\Omega^2$,
which is, according to eq.\,(\ref{Omega}), necessary to give the soliton
the correct isospin. Since the moment of inertia decreases
with increasing constituent quark mass an increasing rotational
frequency is needed
to reproduce the same isospin. Apart from some minor deviations
the Lagrange parameters for nucleons and $\Delta$ isobars differ only in the 
sign. That is why the expectation value $\expv{\vec{T}^{\,2}}_{\rm{hh}}$ lies 
between the corresponding values for nucleon and isobar.
The slight deviations result from the different moments of inertia for 
either particle.

The corrected equation of motion which follows from condition (\ref{eomcorr}) 
is much more involved than eq.\,(\ref{eom1}) since the correction terms 
(\ref{E0trans}--\ref{Ecrank}) depend on the meson profile. 
Therefore, we apply a variation method in order to find a profile function
$\Theta^T_{\rm{corr}}(r)$ which minimizes the corrected soliton energy 
(\ref{Ecorr}). For that aim we represent the profile function by a set of 
numbers $\Theta_i, (i\!=\!1,\ldots,N)$ and $\tilde{m}_\pi$. 
The $\Theta_i\!\equiv\!\Theta(r_i)$ are the values of 
$\Theta(r)$ at $N$ appropriately selected points 
$r_1,\ldots,r_N$ (knots) and $\tilde{m}_\pi$ characterizes the
behavior at $r\!\to\!\infty$.
Between the $r_1$ and $r_N$ 
the function $\Theta(r)$ will be reproduced by means 
of a spline interpolation using the $\Theta_i$'s. 
The asymptotic behavior at small ($r\!<\!r_1$) and large radii 
($r\!>\!r_N$) requires particular care. 
Expanding the profile function of the static soliton in a power 
series one can show that the quadratic term vanishes at small $r$. 
The same is true for the corrected profile function where only the slope is
modified by the translational zero-point correction. 
Rotational corrections do not affect the behavior at small radii.
Therefore we use a linear ansatz at $r\!\le\!r_1$ in agreement with the 
boundary conditions (\ref{ThAsy0})
\be\label{Theasy0}
\Theta(r)\;\stackrel{r\to0}{\longrightarrow}\;
-\pi+\left(\Theta_1+\pi\right)\frac{r}{r_1} 
\;.\ee
The slope is determined by the variation parameter $\Theta_1$.

The behavior at large separations can be determined analytically 
\cite{Meissner-91,Birse-90}. 
Applying the gradient or heat-kernel expansion on the effective action
of the chiral soliton in mean-field approximation 
one can show that the dominating terms agree with the non-linear 
chiral $\sigma$ model with an additional centrifugal force. 
The centrifugal term modifies the asymptotic behavior of the meson profile,
which is proportional to $\rm{e}^{-m_\pi r}$ at large $r$,
in dependence on the rotational frequency $\Omega$. 
In the plane perpendicular to the rotational axis, the pion mass $m_\pi$, 
which characterizes the exponential descent of the meson field, has to be 
replaced by a modified pion mass \cite{Blaizot-88,Post-89,Dorey-94}
\be\label{mpitilde}
\tilde{m}_\pi\,=\,\sqrt{m_\pi^2-\Omega^2}.
\ee
Since $\Omega^2$ is negative for the nucleon and positive for the 
$\Delta$ isobar (\cf fig.\,2) the nucleon is somewhat slimmer ({\it prolate}) 
than the static hedgehog while the isobar is a bit fatter ({\it oblate}). 
Along the cranking axis the soliton is not affected by the rotation. 
Considering this effect in first order we neglect the deviation from 
spherical symmetry and introduce a common effective descent parameter 
$\tilde{m}_\pi$ (effective pion mass) for all directions which is 
considered as an additional variation parameter.
The behavior at $r\ge r_N$ is then determined by
\be\label{Theasy}                                                    
\Theta(r)\;\stackrel{r\to\infty}{\longrightarrow}\;
\Theta_N\,\left(\frac{r_N}{r}\right)^2\frac{1+\tilde{m}_\pi r}
   {1+\tilde{m}_\pi r_N}\,
   \rm{e}^{-\tilde{m}_\pi(r-r_N)} 
\;.\ee
Restricting ourselves to spherically symmetric solitons we average the effect 
of the rotation over all directions of the rotational axis. 
Since only two of the three spatial directions are affected by the rotation 
we expect roughly 
$\tilde{m}_\pi^2\!\approx\!m_\pi^2\!-\!\frac{2}{3}\,\Omega^2$ instead
of eq.\,(\ref{mpitilde}).
Rotational frequencies $\Omega^2\!>m_\pi^2$ lead to negative values of
$\tilde{m}_\pi^2$ in eq.\,(\ref{mpitilde}). The corresponding $\pi$ field 
oscillates perpendicular to the cranking axis and describes the emission of 
pions by the soliton. 
The basis we use for the description of wave functions is defined 
within a finite box \cite{Kahana-84} and hence not appropriate for the 
description of oscillating $\pi$ fields. 
The effective pion mass resulting from the
numerical procedure is close to zero, but real in all cases. In this way
the basis prevents the soliton from emitting pions.\vspace*{-12mm}

\noindent
\bmp{h}{9cm}\hspace*{-4mm}
\mbox{\psfig{file=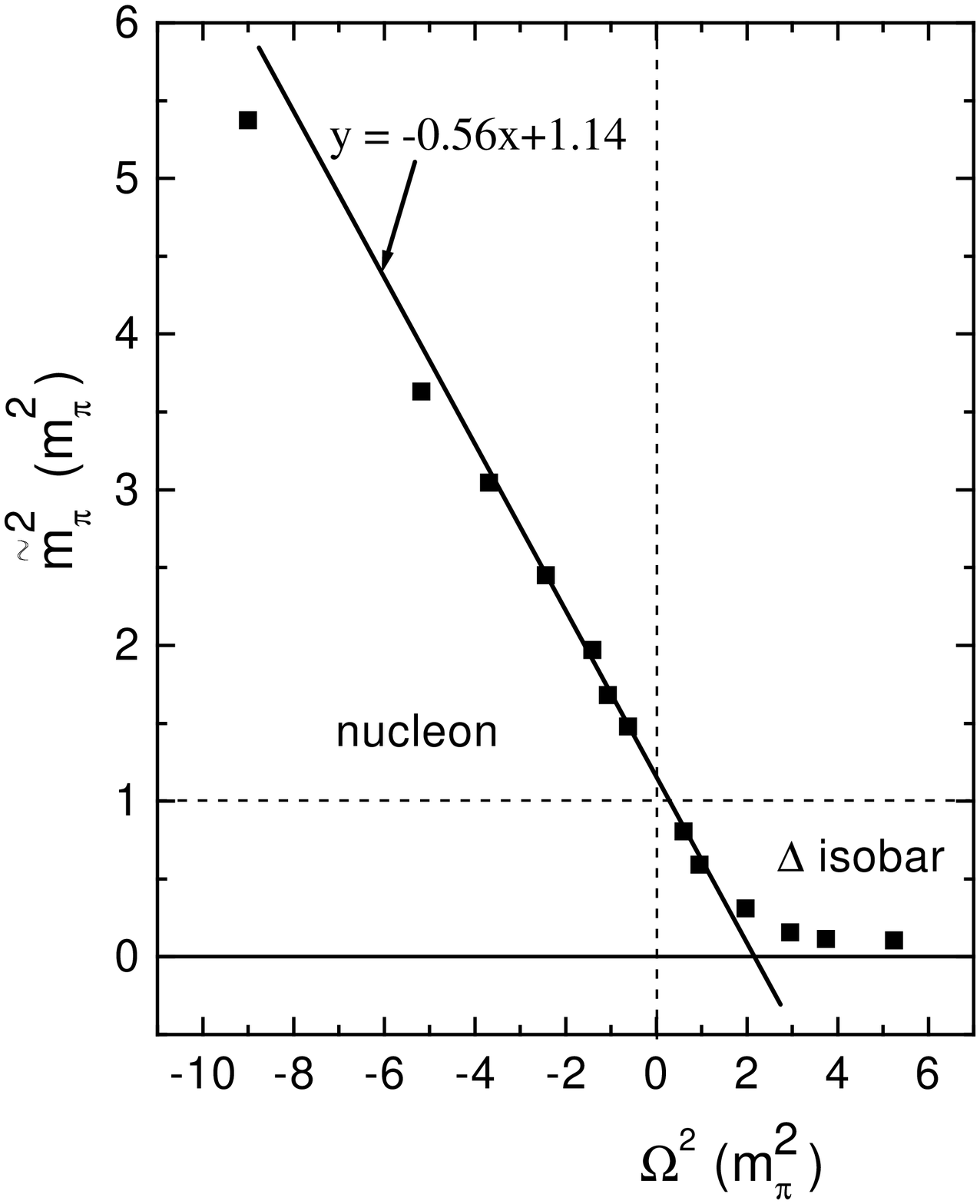,width=9cm,angle=0}}
\emp\hfill\bmp{h}{5.cm}
{\small{\bf Fig.\,3:}\hspace*{1mm}
Square of the effective pion mass $\tilde{m}_\pi$, which characterizes the
asymptotics of the meson field according to \eqn{Theasy}, as obtained from 
minimizing $E^{\rm{N},\Delta}_{\rm{corr}}$ (\ref{Ecorr}), as a function of 
$\Omega^2$.
Different values $\Omega^2$ correspond to different constituent quark masses
according to fig.\,2. 
The line shows a linear fit performed in the region 
\mbox{$-6\le\left(\Omega/m_\pi\right)^2\le 1$}.\baselineskip10pt}
\emp

Fig.\,3 displays the numerically determined effective pion masses as a function
of $\Omega^2$. A fit in the region $-6m_\pi^2\!\le\!\Omega^2\!\le\!m_\pi^2$ 
yields $\tilde{m}_\pi^2\!\approx\!1.14\,m_\pi^2\!-\!0.56\,\Omega^2$ in fair 
agreement with our suggestion above. 
Interpolating to $\Omega^2\!=\!$ 0 we get an effective pion mass 
which deviates from the pion rest mass by 7 percent. 
A similar deviation (5--10 percent) from the pion rest mass was obtained 
in a numerical variation without any rotational corrections 
(only translational corrections or no corrections at all).  
This demonstrates the accuracy of our procedure. 
The obvious deviation from the straight line at larger positive values of 
$\Omega^2$ in fig.\,3 reveals the limitation to descending pion field in our
numerical calculation.

Representing the profile function $\Theta(r)$ by the set of parameters 
$\Theta_i, (i\!=\!1,\ldots,N)$ and $\tilde{m}_\pi$ the effective 
energies (\ref{Emf}) and (\ref{Ecorr}) are ordinary functions of $N\!+\!1$ 
variables which can be minimized by standard methods.
We tested the method by minimizing the static mean-field energy where the
self-consistent profile $\Theta(r)$ can alternatively be determined in the 
traditional manner by solving the equation of motion (\ref{eom}) iteratively. 
Here we found that 7 knots are sufficient for an accurate reproduction 
of the meson profile. Moreover, the optimal position of the knots was  
determined. Details can be found in ref.\,\cite{Schleif-94}.
Since the corrected profiles do not substantially deviate
from the uncorrected ones we used the same knots for the
representation of $\Theta_{\rm{corr}}^T(r)$. \vspace*{-5mm}

\noindent
\bmp{h}{9cm}\hspace*{-4mm}
\mbox{\psfig{file=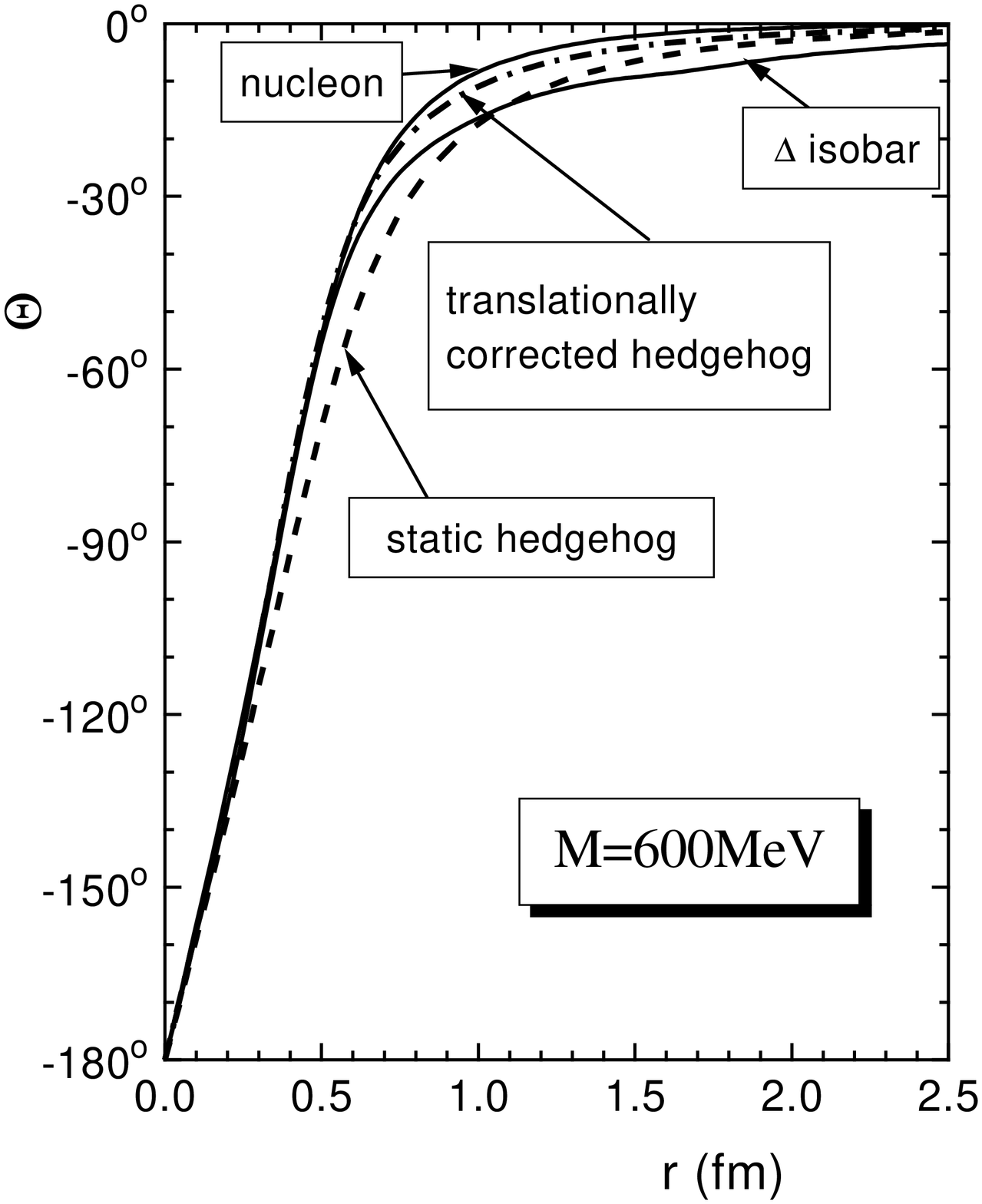,width=9cm,angle=0}}
\emp\hfill\bmp{h}{4.5cm}
{\small{\bf Fig.\,4:}\hspace*{1mm}
Profile functions $\Theta(r)$ for the static hedgehog soliton 
({\em broken line}) in \mbox{comparison with the} corrected profiles 
for nucleon \mbox{and $\Delta$ isobar ({\em full lines})} cal\-culated for the 
con\-sti\-tu\-ent quark mass $M\!=\!600\MeV$. 
Additionally the self-consistent profile for a 
\mbox{translationally corrected} 
\mbox{soliton without rotational} cor\-rec\-tion is shown\\
({\em dashed-dotted line}).\baselineskip10pt}
\emp

Fig.\,4 illustrates the general features of the modification in the profile 
function caused by the correction terms. We selected the relatively large 
constituent quark mass of $600\MeV$ in order to get a pronounced effect. 
Translational corrections modify the behavior of the meson profile at small
radii while rotational corrections affect the asymptotics at large
radii. 
The first ones increase the slope of the internal linear part and
make the soliton smaller. 
The size of the mean field and of the corresponding quark distribution is a 
result of the balance between the diverging Fermi motion of the quarks and 
the attraction between them. 
Subtracting the translational zero-point energy the balance is disturbed 
since the Fermi motion is reduced and the attraction dominates.
The soliton shrinks until the increasing Fermi motion balances the 
attraction.

Rotational corrections are different in sign for nucleon and $\Delta$ isobar
and manifest themselves at large $r$. They act like a centrifugal force,
which is negative for the nucleon and positive for the $\Delta$ isobar. 
In accordance with eq.\,(\ref{mpitilde}) and fig.\,2 the exponential descent 
is larger for nucleons and smaller for $\Delta$ isobars. 
Fig.\,4 shows that rotational corrections modifies the meson field already at 
intermediate distances from the center of the soliton 
($r\gapp0.6\fm$). \vspace*{-10mm}

\centerline{\psfig{file=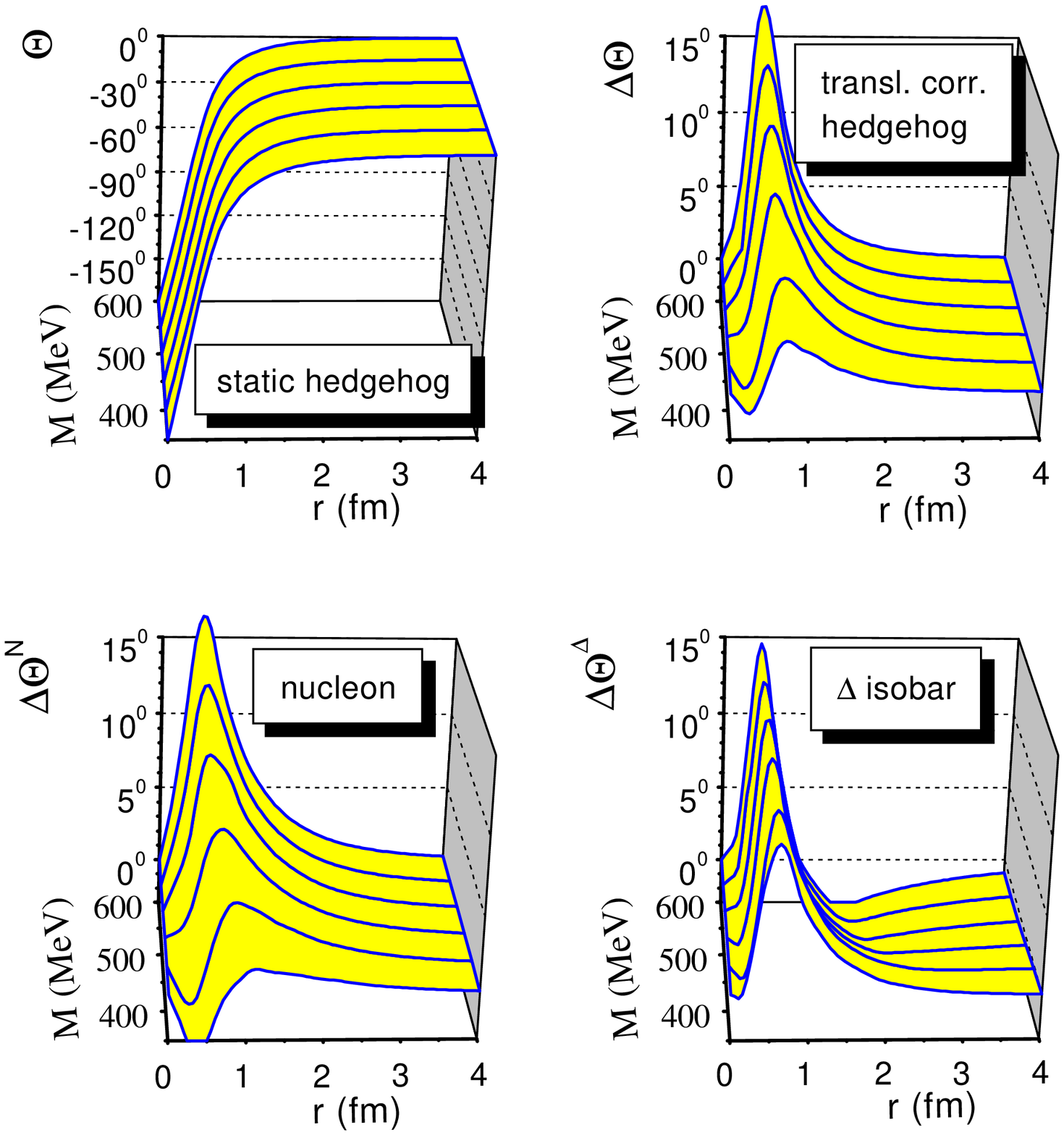,width=12cm,angle=0}}\vspace*{-25mm}

\noindent{\small{\bf Fig.\,5:}\hspace*{1mm}
{\it Upper left figure}: 
Profile function $\Theta(r)$ for the static soliton in the mass region
$350\MeV\le M\le600\MeV$;
{\it Upper right figure}: 
Deviation $\Delta\Theta(r)$ of the translationally corrected profile 
from the static one;
{\it Lower figures}:
Deviations $\Delta\Theta^{\rm{N},\Delta}(r)$ of the translationally and 
rotationally corrected meson profiles for nucleon and $\Delta$ isobar, 
respectively.\baselineskip10pt}
\vspace*{5mm}

Fig.\,5 displays the self-consistent meson profiles and the deviations
of the corrected profiles from the static ones in the whole mass region. 
As already shown in ref.\,\cite{Wuensch-94} the shape of the static profile is 
practically independent of the constituent quark mass (upper left figure). 
The distortion of the static meson profile by the correction terms 
turns out to be more significant than its variation in dependence 
on $M$. 
The effect of the corrections grows in dependence on the
constituent quark mass $M$. Heavier constituent quarks are stronger bound
and produce smaller solitons. Their center-of-mass energy is larger, while
the moment of inertia is smaller. Both factors increase the correction term
and their influence on the meson profile.

\section{Expectation values of the soliton \\
calculated with corrected meson profiles}

It is the aim of this paper to evaluate the influence of translational and 
rotational corrections of the self-consistent meson fields and to estimate 
their effect on expectation values of solitonic observables. 
For that reason we compare expectation values of 
observables which are calculated with corrected and uncorrected
meson profiles, respectively.
In the present paper we will not consider pushing and cranking 
correction to the observables themselves.
A full treatment of symmetry restoration would have to include these 
corrections as well. In several cases they have turned out to be rather 
important \cite{Wakamatsu-93,Christov-94} and therefore must be taken care of 
before the results are compared with experimental data.
This will be the subject of a forthcoming analysis.

Generally it can be said that
the effect coming from the difference between uncorrected
and corrected meson profiles
grows with increasing constituent quark mass $M$ 
in accordance with the increasing size of the correction terms (\cf fig.\,1). 
At small constituent quark masses $M$ the valence quarks 
give the dominating contribution to the expectation value in most cases. 
Increasing the quark mass the Dirac sea gets more and more polarized 
and its contribution gets larger. 
Covering the mass region $300\MeV\!\le\!M\!\le\!600\MeV$ we are able to 
control the ratio between the contributions resulting from valence and 
sea quarks. 

Fig.\,6 illustrates the energy of the soliton and its components.
The total corrected energies for nucleons and $\Delta$ isobars
are shown in the upper part. 
The experimental nucleon mass is reached at $M\!\approx\!350\MeV$, but the
predicted mass of the isobar is remarkably smaller than the experimental value
in the whole mass region. 
Pushing and cranking corrections in the meson field decrease 
the mass furthermore.
The difference between full and broken lines
indicates the gain of energy which is obtained if one minimizes the corrected 
soliton energy instead of its static value.
Here we compare the corrected total soliton energy (\ref{Ecorr}) calculated
for the corrected profiles $\Theta_{\rm{corr}}^T$ with the same energy 
calculated for the static profile $\Theta$. 
While the total energies do not differ very much from each other 
valence and sea quark energies are rather different in both cases
(central and lower part of fig.\,6). 
For the static soliton we have a dominating valence contribution at small 
constituent quark masses ($M\lapp420\MeV$). 
In the energy of the corrected soliton, the valence quarks dominate up to 
larger masses ($M\lapp500\MeV$). 
The correction terms even prevent the valence quark energy from getting 
negative what happens at $M\!\approx\!750\MeV$ in the uncorrected case 
\cite{Wuensch-94}. 
The field energy (\ref{Em}) of the mesons is strongly affected by the 
correction (up to a factor of 2). However, its contribution to the 
total soliton energy is marginal due to the restriction to the chiral circle.
\vspace*{-5mm}

\noindent
\bmp{h}{9.5cm}\hspace*{-2mm}
\mbox{\psfig{file=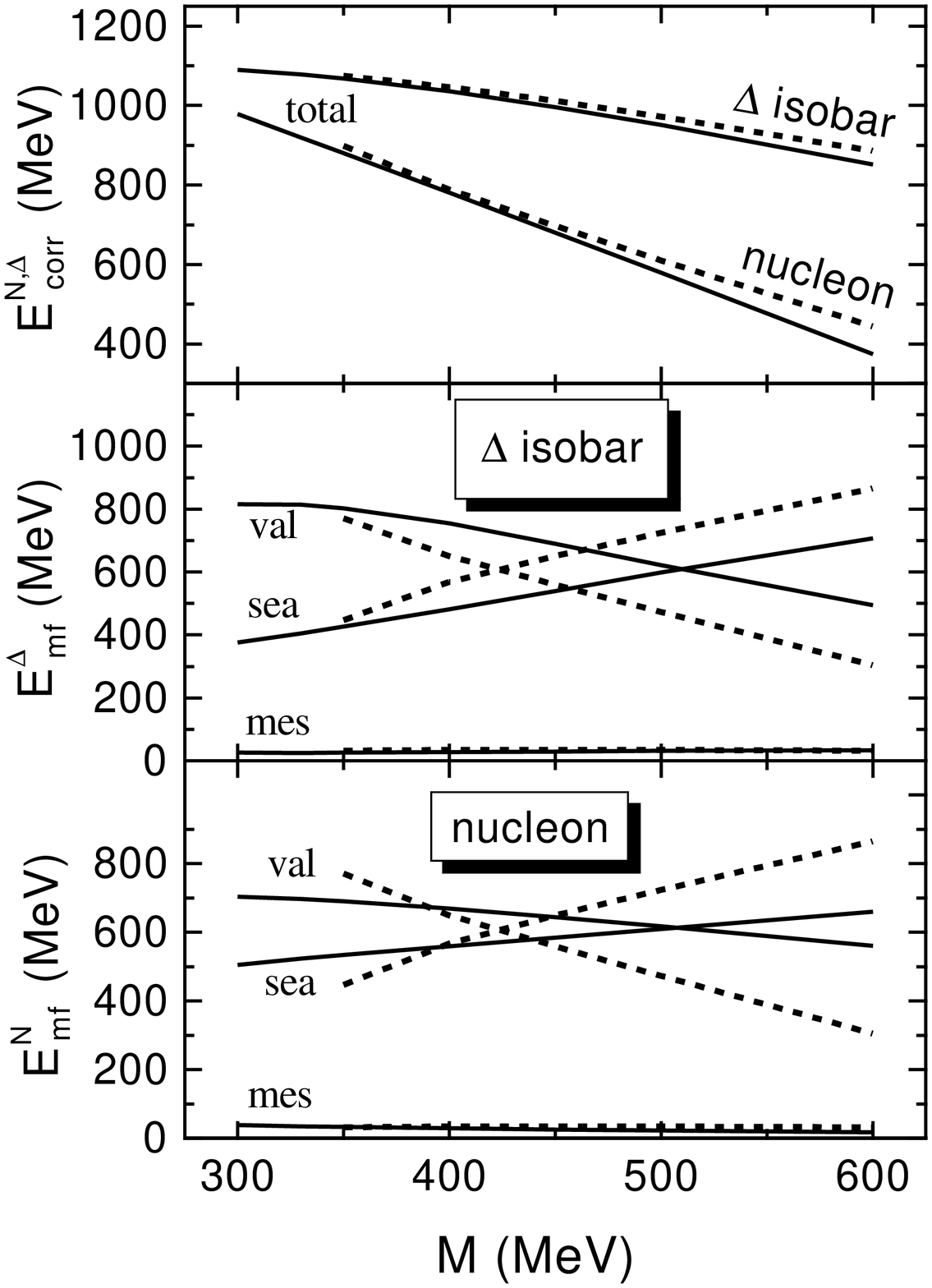,width=10cm,angle=0}}
\emp\hfill\bmp{h}{4.5cm}
{\small{\bf Fig.\,6:}\hspace*{1mm}
{\em Upper part}: 
Total corrected energies $E^{\rm{N},\Delta}_{\rm{corr}}$ for nucleon and 
$\Delta$ isobar calculated for the static meson profile $\Theta$
({\em broken lines}) in comparison to the same energy calculated for the 
corrected profiles $\Theta^{\rm{N},\Delta}_{\rm{corr}}$ 
({\em full lines}).

{\em Central and lower part}: Contributions of the valence (val) and 
sea quarks (sea), and of the mesonic field (mes) to the total soliton 
energy $E_{\rm{mf}}$ calculated for the static meson profile 
({\em broken lines}) and for the corrected profiles 
({\em full lines}) as a function of the constituent quark mass $M$.
\baselineskip10pt}
\emp

The effect of the corrections on an expectation value depends
crucially on the relation between valence and sea-quark contributions.
If the valence quarks, which are usually well bound inside the soliton, 
give the dominating contribution, the shrinking of the meson profile 
in the internal region (see fig.\,4) is the most important effect. 
It is caused by the center-of-mass correction and is independent 
of the isospin quantum number. 
Rotational corrections have almost no effect on the valence quarks.
Sea quarks are either less bound or unbound and hence experience the action
of rotational corrections which are opposite for nucleons 
and $\Delta$ isobars. Considering the sea contribution of an expectation value
in dependence on the radius $r$ one notices that contributions from the 
internal part of the soliton may cancel out contributions 
from external layers. 
This is not the case for the valence part since the corresponding 
wave function has no node.
Modifications in the meson profile influence the complicated interplay between 
positive and negative contributions and may affect the total sea contribution 
considerably. 
As a result sign and size of the effect depend strongly on 
the observable. 
In the following we give some examples in order to illustrate what might 
happen.\vspace*{-5mm}
 
\noindent
\bmp{h}{9.5cm}\hspace*{-2mm}
\mbox{\psfig{file=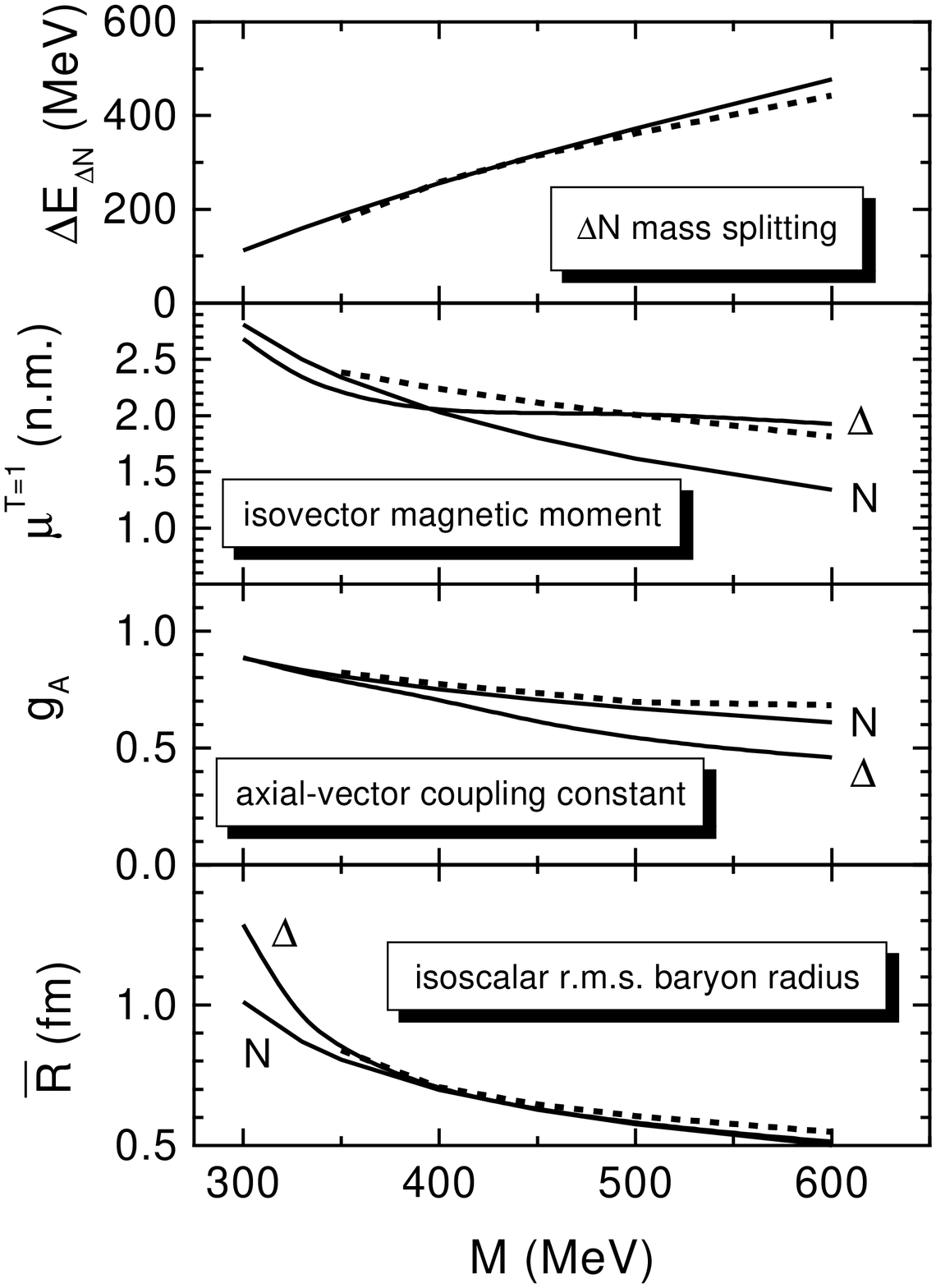,width=10cm,angle=0}}
\emp\hfill\bmp{h}{4.2cm}
{\small{\bf Fig.\,7:}\hspace*{1mm}
$\Delta$-nucleon mass splitting $\Delta E_{\Delta \rm{N}}$, 
isovector magnetic moment \linebreak 
$\mu^{T=1}\equiv\mu_{\rm{p}}\!-\!\mu_{\rm{n}}$,
axial-vector coupling strength $g_{\rm{A}}$ and isoscalar root 
mean-squared (r.\,m.\,s.) baryon radius $\bar{R}$ calculated for the 
translationally and rotationally corrected nucleon (N) 
and $\Delta$ profiles ({\em full lines}) in comparison to the same 
quantities for the static soliton ({\em broken lines}), as a function of 
the constituent quark mass $M$.\baselineskip10pt}
\emp

The (isoscalar) r.\,m.\,s.~radius $\bar{R}$ of the quark distribution 
(lower part of fig.\,7) is dominated by the valence quarks by nearly 
95 percent. 
Hence nucleon and $\Delta$ isobar have almost the same radius. 
The reduction of the radius with respect to the static hedgehog by a few 
percent is in quantitative agreement with the steeper behavior of the 
corrected meson profile in the central region of the soliton (\cf figs.\,4,5).
For very small constituent quark masses the valence quarks are so weakly 
bound that they may reach larger separations from the center. 
Here they are exposed to the rotational corrections which lead to a 
considerable shrinking for the nucleon and a remarkable swelling for the 
$\Delta$ isobar with respect to the static hedgehog. The r.\,m.\,s.~radius 
is particularly sensitive to the quark distribution at large radii 
since the quark density is weighted by a factor $r^4$.

Now let us investigate the axial-vector coupling constant $g_{\rm{A}}$ 
of the nucleon (fig.\,7).  
In order to study the pure influence of the corrected meson field we 
consider the expectation value of the same operator \cite{Wuensch-94} 
without any rotational corrections. 
This uncorrected observable accounts for roughly 65 percent of the 
experimental value of $g_{\rm{A}}$ \cite{Wakamatsu-93,Christov-94}. Apart from 
rotational corrections the coupling constant for the $\Delta$ isobar
differs from the nucleon by an additional factor which results from the 
different spin-isospin structure of both particles. The figure shows the 
$g_{\rm{A}}$ of the proton calculated for meson profiles which have been 
corrected for both nucleons and $\Delta$ isobars.
Here the dominance of the valence quarks is less pronounced as in the case of 
$\bar{R}$ and the different asymptotic behavior of nucleon and $\Delta$
profiles results in slightly different expectation values. 
Roughly spoken translational and rotational corrections cancel each other 
for the nucleon while they add up (negatively) for the $\Delta$ isobar.   

In the isovector magnetic moment of the nucleon 
($\mu^{T=1}_{\rm{N}}\!\equiv\!\mu_{\rm{p}}\!-\!\mu_{\rm{n}}$), 
contributions from larger radii 
are favored by a factor $r^3$ \cite{Christov-96,Christov-94}.
Sea contributions amount to 20\dots30 percent of the total value.  
This leads to a noticeable difference between the expectation values for 
nucleon and $\Delta$ profiles. Here, the correction is negative for 
the nucleon and almost cancels out for the $\Delta$ isobar.

In the last example we consider the delta-nucleon mass splitting 
$\Delta E_{\rm{\Delta N}}$.
In the case of the static soliton the splitting results solely from the 
different cranking energies of both particles and is uniquely determined by the
moment of inertia $I$ 
\be\label{DNM}
\Delta E_{\rm{\Delta N}}\,=\,\frac{3}{2I}.
\ee
Using the corrected meson profiles, which are slightly different for nucleons
and $\Delta$ isobars, the self-consistent values of static energy 
(\ref{Emf}), translational zero-point energy (\ref{E0trans}) 
and of the moment of inertia are different as well. 
Now the mass difference is given by  
\be\label{DNMSCORR}
\Delta E_{\rm{\Delta N}}^{\rm{corr}}\,=\,
\left[E^\Delta_{\rm{mf}}-E^{\rm{N}}_{\rm{mf}}\right]-
\left[E_{\rm{trans}}^{o,\Delta}-E_{\rm{trans}}^{o,\rm{N}}\right]+
\frac{3}{4}\left[\frac{1}{I^\Delta}+\frac{1}{I^{\rm{N}}}\right]
\ee
where the additional upper index $N,\Delta$ indicates the corresponding 
particle. In the upper part of fig.\,7 we see that the simple expression 
(\ref{DNM}) with the static moment of inertia agrees with the
more complicated expression (\ref{DNMSCORR}) for corrected profiles very well. 
The analysis of the various terms in eq.\,(\ref{DNMSCORR}) 
shows that the first two differences are rather small ($\approx\!10\MeV$).
In spite of slightly different moments (\cf fig.\,2) of nucleon and $\Delta$
isobar the last term is nearly identical with the result of eq.\,(\ref{DNM}).
That means that the harmonic average defined by
\be\label{Iav}
\frac{1}{I^{\rm{av}}}\,=\,
\frac{1}{2}\left[\frac{1}{I^\Delta}+\frac{1}{I^{\rm{N}}}\right]
\ee
agrees with the moment of the static hedgehog soliton (\cf fig.\,2).
The mass splitting provides an example of an almost complete canceling of
meson-field corrections.

The magnitude of the evaluated effects does not exceed a level which 
describes the reliability of the model in particular in the physically 
relevant region of small constituent quark masses. 
The situation may be different if one considers observables with
oscillating radial dependence, \eg formfactors. Here a minor modification in
the meson profile may have a larger effect.

\section{Conclusions}

We have evaluated pushing and cranking corrections on solitonic meson fields 
of the bosonized NJL model in one-loop approximation and studied their effect
on the quark distribution and other observables of the soliton.
The corrected fields have been obtained by minimizing an energy functional 
which differs from the mean-field energy by several correction terms. 
These terms have been introduced to remove the energy of spurious
translational and rotational modes and to equip the soliton with the 
quantum numbers of nucleon or $\Delta$ isobar on a semi-classical level.
The meson fields are restricted to the chiral circle and to 
hedgehog configurations with a modified asymptotic behavior.

We have studied solitons and their expectation values in the region 
$300\MeV\!\le\!M\!\le\!600\MeV$ 
of the constituent quark masses $M$.  
The results illustrate the response of the meson field to the corrections
and quantify their effect on expectation values. 
Despite the big correction terms (up to 50 percent of the total soliton
energy) meson and quark fields are only moderately affected. 
The (approximate) restoration of translational symmetry increases 
the slope of the meson profile in the central region and makes the soliton 
smaller. This effect prevents the valence quarks from diving into the 
negative-energy region up to very large constituent quark masses.
It stabilizes the soliton by reducing the kinetic energy of the quarks.  
As a result, a stable soliton exists at quark masses below 350\MeV 
(up to $\approx\!300\MeV$) while the uncorrected soliton is heavier 
than 3 free quarks and decays, in this mass region.
Rotational corrections affect the asymptotic behavior of the fields at large 
radii. They depend on the 
spin and isospin quantum numbers and therefore have different effects 
on nucleon and $\Delta$ isobar.
Both translational and rotational corrections grow with increasing 
constituent quark mass. 

Expectation values of the soliton are affected by the corrections to
various extents.
The isoscalar r.\,m.\,s.~radius of the nucleon is reduced by only a few 
percent ($\approx\,$3--5 percent). 
Larger changes can be observed if one considers valence and sea quark
contributions separately.  
Axial-vector coupling strength and isovector magnetic moment are modified
up to 40 percent.
In the physically relevant region of small constituent quark masses 
($350\MeV\lapp M\lapp450\MeV$), the corrections do not exceed 20 percent. 
The $\Delta$-nucleon mass splitting is practically not affected by the 
corrections despite the slightly different moments of inertia calculated for
nucleon and $\Delta$ isobar.

The experimental values of r.\,m.\,s.~radius and $\Delta$N mass splitting
are well reproduced for light constituent quark masses around 400\MeV. Here,
the effect of translational and rotational corrections to the mean field is
small and does not exceed the level of accuracy of the model and its numerical
treatment. The quantities $g_{\rm{A}}$ and $\mu^{T=1}$ are stronger affected
by the corrections. 
Up to now the theoretical values underestimate the 
experimental ones significantly. 
The calculated values are even smaller with the corrections than without.
However it must be noted that our treatment is incomplete, because in order to
be able to compare with the experiment one has to take into account that there are
also rotational corrections to the corresponding 
operators, which have turned out be quite large \cite{Christov-96}. \\

\noindent{\sc Acknowledgement}

\noindent 
The authors wish to acknowledge stimulating discussions with K.\,Goeke,
\linebreak 
H.\,Reinhardt, R.\,Alkofer, H.\,Weigel, J.\,Berger and Chr.\,V.\,Christov.
The paper was supported by the Bundesministerium f\"ur Forschung      
und Technologie (contract 06 DR 666).  

\renewcommand{\thesection}{\arabic{section}}
\renewcommand{\theequation}{\thesection.\arabic{equation}}
\renewcommand{\labelenumii}   {\arabic{enumi}.\arabic{enumii}}
\setcounter{equation}{0}

%\newpage

\end{document}